# Influence of Molecular Solvation on the Conformation of Star Polymers

*Xin Li[†], Lionel Porcar[‡], Luis E. Sánchez-Diáz[†], Changwoo Do[†], Yun Liu[§,#], Tae-Hwan Kim[∥], Gregory S. Smith[†], William A. Hamilton[∇], Kunlun Hong[⊥,\*], and Wei-Ren Chen[†,\*]*

[†]Biology and Soft Matter Division, Oak Ridge National Laboratory, Oak Ridge, Tennessee 37831, USA

[‡]Institut Laue-Langevin, B.P. 156, F-38042 Grenoble CEDEX 9, France

[§]The NIST Center for Neutron Research, National Institute of Standards and Technology, Gaithersburg, Maryland 20899-6100, USA

[#]Department of Chemical Engineering, University of Delaware, Newark, Delaware 19716, USA

[∥]Research Reactor Utilization Department, Korea Atomic Energy Research Institute, Daejeon 305-353, Korea

[∇]Instrument and Source Division, Oak Ridge National Laboratory, Oak Ridge, Tennessee 37831, USA

[⊥]Center for Nanophase Materials Sciences, Oak Ridge National Laboratory, Oak Ridge, Tennessee 37831, USA.

Corresponding author's email address: chenw@ornl.gov; hongkq@ornl.gov

ABSTRACT: We have used neutron scattering to investigate the influence of concentration on the conformation of a star polymer. By varying the contrast between the solvent and isotopically labeled stars, we obtain the distributions of polymer and solvent within a star polymer from analysis of scattering data. A correlation between the local desolvation and the inward folding of star branches is discovered. From the perspective of thermodynamics, we find an analogy between the mechanism of polymer localization driven by solvent depletion and that of the hydrophobic collapse of polymers in solutions.

Star polymers are synthetic macromolecules consisting of linear polymer branches emanating from the molecular center.[1-4] They are characterized by the structural features of both linear polymers and hard colloids. There is wide interest in their properties because of their structural novelty as well as their role in a variety of industrial applications.[1,5] It is now an important family of soft matter.[4]

There has been much interest in understanding the conformational properties of an isolated star including the density profile, overall size, and scattering functions.[1-4, 6-23] For each single polymer branch, the steric hindrance decreases steadily with the distance from the molecular center.[1] As a result the intra-star polymer density shows a non-uniform distribution which decreases from the center towards the periphery. Scaling criteria have been developed to categorize the intra-star spatial regions characterized by different conformational features.[1,4,10]

Considerable effort has also been devoted to investigating the manner in which the inter-star interactions affect the individual star conformations due to their flexible molecular architecture. Particularly the conformational dependence of a star on its concentration has been examined by scattering experiments.[17,19] Results suggest that while the global size of a star, in terms of the radius of gyration ($R_G$), essentially exhibits no concentration dependence when the concentration, $c$, is less than the overlap concentration $c^*$, $R_G$ indeed decreases progressively when $c$ is raised above $c^*$. However, what $R_G$ reflects is the integrated information of the intra-star density profile.[24] The detailed conformational variation, which involves the relocation of the polymer components and invasive solvent within a star, cannot be directly obtained from this coarse-grained information. At present it is still difficult to computationally simulate this concentration effect in which solvent and polymer are incorporated explicitly. Therefore for star polymers the microscopic mechanism of concentration-driven conformational evolution remains to be explored. This challenge provides the main motivation for this study.

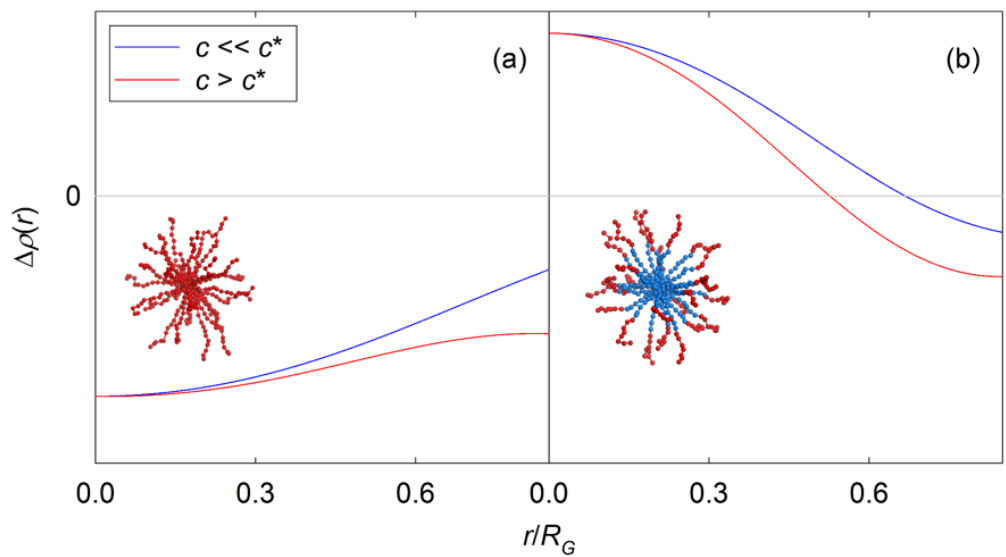

**Figure 1.** Schematic representation of the dependence of scattering contrast on concentration for (a) fully protonated star and (b) star with deuterated interior. $\Delta\rho(r)$ is the excess scattering length density (SLD) of a star in a deuterated solvent. The protonated and deuterated components of a star are represented by red and blue beads respectively. The straight gray lines mark the SLD of deuterated solvent.

Small angle scattering techniques, including both neutron (SANS) and x-ray (SAXS), provide experimental tools to obtain the conformation of stars in solution because of their accessible nanoscale spatial resolution. When $c > c^*$, the local polymer density around the star periphery increases because of the interpenetrating polymer arms. As a result, if a star is fully protonated, the intra-star distribution of excess scattering length density (SLD) $\Delta\rho(r)$ of a strained star be-comes more homogeneous (red curve) than that of its isolated state (blue curve). Therefore, for both SANS and SAXS, the inter-star interpenetration inevitably leads to a significant loss of scattering contrast. This concentration effect on the scattering behavior is conceptualized in Figure 1(a). However, this intrinsic constraint can be bypassed by investigating a star with interior and periphery bearing the opposite isotopic label using SANS. Because the bound scattering lengths of a proton and a deuteron have opposite signs, the increase in the polymer density at the periphery in fact accentuates the SLD difference between the interior and that of periphery as demonstrated in Figure 1(b).

Therefore, in this report we investigate isotopically labeled polystyrene (PS) stars dissolved in a good solvent, tetrahydrofuran (THF) using SANS. THF solutions of fully protonated stars are investigated as a reference system. Both PS stars were synthesized at the Center for Nanophase Materials Sciences (CNMS), Oak Ridge National Laboratory (ORNL). The molecular weight determined from light scattering was found to be 538 k. The number of arms was found to be around 26. In the labeled stars, the protons residing in the interior of a star (around 60% of overall constituent protons according to SANS data analysis) are isotopically replaced by deuterons. The $c^*$ of our system is estimated to be 22.5 wt% according to reference 16. More details of material synthesis are given in the Supporting Information (SI). The effect of isotopic labeling on the scattering signature of a star is clearly confirmed by the evolution of SANS intensity $I(Q)$ shown in the SI. Upon increasing the concentration, while an expected decrease in the $I(Q)$ for the fully protonated PS star is seen, the intensity of $I(Q)$ for the PS star with a deuterated core indeed increases continuously.

Since the flexible open architecture of the stars allow significant solvent penetration, we consider how this might affect the star conformation. For a SANS experiment the conformational information of a star is reflected by the zero-angle scattering $P(0)$,[24]

$$P(0) = nb_{total}^2 \tag{1}$$

where $n$ is the number density of stars in THF and $b_{total}$ is the sum of the bound scattering lengths of the constituent atoms of a single star. $b_{total}$ can be explicitly expressed as

$$b_{total} = 4\pi \int_0^\infty \rho(r) r^2 dr = 4\pi \int_0^\infty [\rho_{polymer}(r) + \rho_{THF}(v_{THF}(r) - 1)] r^2 dr \tag{2}$$

where $\rho_{polymer}(r)$ and $v_{THF}(r)$ are the SLD distributions of polymer and volume fraction of invasive solvent within a star, respectively. $\rho_{THF}$ is the SLD of THF. Combining eqns. (1) and (2), $P(0)$ is found to take the following expression

$$P(0) = n(\Delta b_s \gamma - b_0)^2 \tag{3}$$

where $\Delta b_s$ is the product of the SLD difference between the deuterated and protonated THF, and the volume of polymeric components. $b_o$ is the scattering power of a star immersed in fully protonated THF. $\gamma$ is the solvent deuterium-hydrogen (D/H) volume fraction. By changing $\gamma$, one can alter $P(0)$ without affecting the star conformation. Therefore, this contrasting technique provides a probe to assess the structural inhomogeneity within a star. Figure 2 presents the experimental $P(0)$ for fully protonated and partially deuterated PS stars. Measurements of $P(0)$ obtained at different values of $\gamma$ are normalized by that obtained in $D_2O$. The isotopic labeling is seen to influence the evolution of $P(0)$ significantly. By taking the derivative of eqn. (3) one finds

$$\frac{dP(0)}{d\gamma} = 2n\Delta b_s(\Delta b_s \gamma - b_0) \qquad (4)$$

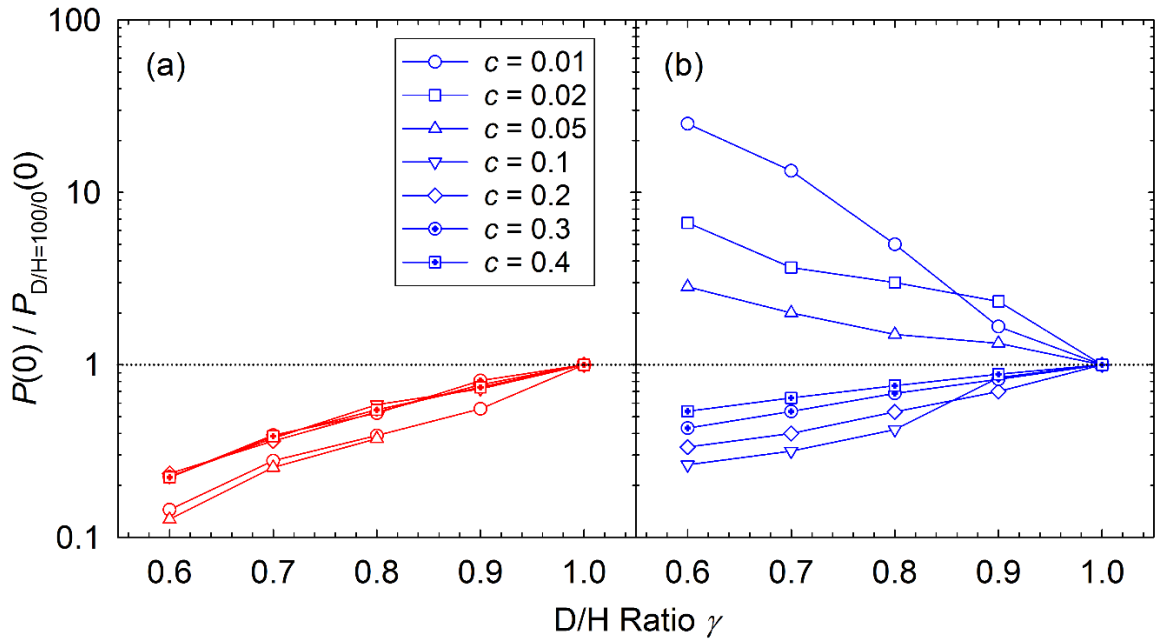

**Figure 2.** The normalized zero-angle scattering intensities of (a) fully protonated PS star and (b) deuterated-core PS star solutions in THF as a function of star weight fraction c at different D/H ratios $X$.

Since the bound scattering length of hydrogen is -3.8 fm, $b_o$ for a fully protonated star is a negligible positive number which is more than an order of magnitude less than $\Delta b_s \gamma$ within the probed range of $\gamma$. Therefore from eqn. (3) it is seen that the corresponding $P(0)$ increases monotonically with increasing $\gamma$.

Moreover, as shown in Figure 2(a), after normalization, the evolution of $P(0)$ essentially exhibits no concentration dependence when $c > 0.1$. This observation indicates that the possible conformational evolution is masked by the lack of scattering contrast. Meanwhile, since the bound scattering length of the deuteron is 6.5 fm, for partially deuterated star, $b_o$ and $\Delta b_s \gamma$ are of comparable magnitude. From eqn. (3), this means that a small conformational change can be amplified by SANS combining isotopic substitution of the star and contrast variation of the solvent. Indeed, upon increasing $c$, the occurrence of conformational variation and its close correlation with solvent penetration is qualitatively demonstrated by the characteristic variation of $\frac{dP(0)}{d\gamma}$ given in Figure 2(b).

Having demonstrated the critical role solvent penetration plays in determining the star conformation, we develop a model scattering function to complement the experiment. Based on the aforementioned dense-core profile, the intra-star solvent and polymer distributions are explicitly parameterized in the form factor $P(Q)$ of a star. The SANS absolute intensity $I(Q)$ of concentrated star solutions not only contains the conformational information of a single star in the form factor $P(Q)$, but also that of the inter-star spatial arrangement in the structure factor $S(Q)$. Since $I(Q)$ effectively can be factorized as the product of $P(Q)$ and $S(Q)$, the former can be effectively compartmentalized by normalizing the $I(Q)$ obtained at a certain $\gamma$ to that obtained at $\gamma = 1$. One example of normalized $I(Q)$ is given in Figure 3. It contains the conformational evolution of a single star with concentration. For fully protonated star solutions, the normalized $I(Q)$ (Figure 3(a)) is rather featureless. On the contrary, upon increasing $c$ the normalized $I(Q)$ for the partially deuterated star (Figure 3(b)) is seen to vary characteristically. The additional scattering features of Figure 3(b), such as the locations of minima, originate from the core-shell SLD profile due to the isotopic substitution. They prove critical for extracting the conformational parameters and allow the determination of the distributions of solvent and polymer within a star separately and accurately. More details of model fitting are given in SI.

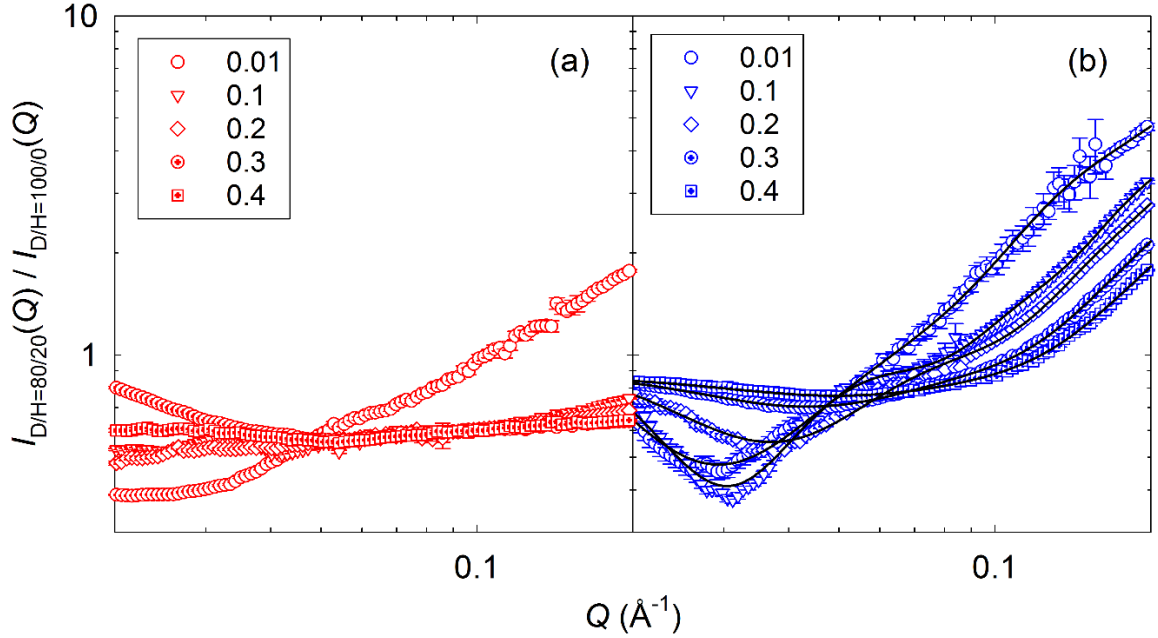

**Figure 3.** The normalized SANS absolute intensity ($I_{80/20}(Q)/I_{100/0}(Q)$) of (a) a fully protonated PS star and (b) a PS star with deuterated core in THF solutions as a function of star weight fraction $c$. The black curves in panel (b) are the results of the theoretical model.

We summarize the main results of the star conformations obtained from SANS model fitting in Figure 4. Figure 4(a) gives the evolution of $R_G$ as a function of $c$. $R_G$ is seen to decrease continuously with increasing star concentration beyond $c^*$. A similar observation has also been reported previously in a different system.[17,19] To provide the physical origin of this size shrinkage, we evaluate the conformational evolution via scrutinizing the structural characteristics within the spatial resolution. In Figure 4(b) we present the radial density distributions of THF (red curves) and PS (blue curves) within a star deduced from our SANS fitting. Upon increasing $c$, the volume fraction of THF is seen to decrease continuously in the core region of 50 Å. Meanwhile, within the same region a progressive increase of PS density is also seen. From the ratio of the volume fraction of THF to that of PS presented in Figure 4(c), denoted as $S(r)$, it is clearly seen that the average THF density is seen to decrease along the radial direction of a PS star. These two observations are correlated based on the following argument. From the perspective of interfacial free energy,[25-26] the free energy of the system is related to the solvent accessible surface area (SASA) of PS branches. Despite the difference in dimensionality, one can draw a direct connection between $S(r)$ and SASA. Because THF is a very good solvent for PS at room temperature, for an isolated PS star its branches are expected to be extended in THF. At high concentrations, the diminishing number density of THF in the vicinity of PS provides the driving force for the PS star to reduce its stretched branch entropically into a random coil-like conformation with lower free energy. As a result, an inward relocation of PS branches is triggered by the molecular desolvation. Therefore, the size shrinkage is a reflection of this conformational evolution occurring at the microscopic level. Meanwhile, upon increasing the concentration of PS stars in the solution, the degree of inter-star interpenetration and congestion are expected to increase progressively. The enhancing physical contact among the PS components consequently causes the exodus of THF solvent. Therefore, the contribution of excluded volume effect to our experimental observation is also obvious. It is instructive to point out that the temperature-triggered hydrophobic collapse of polymers in aqueous solutions has also been attributed to the variation of local hydration level.[27-28] Our experimental finding suggests the relevance of solvent fluctuation in understanding the equilibrium configuration of general soft matter systems.

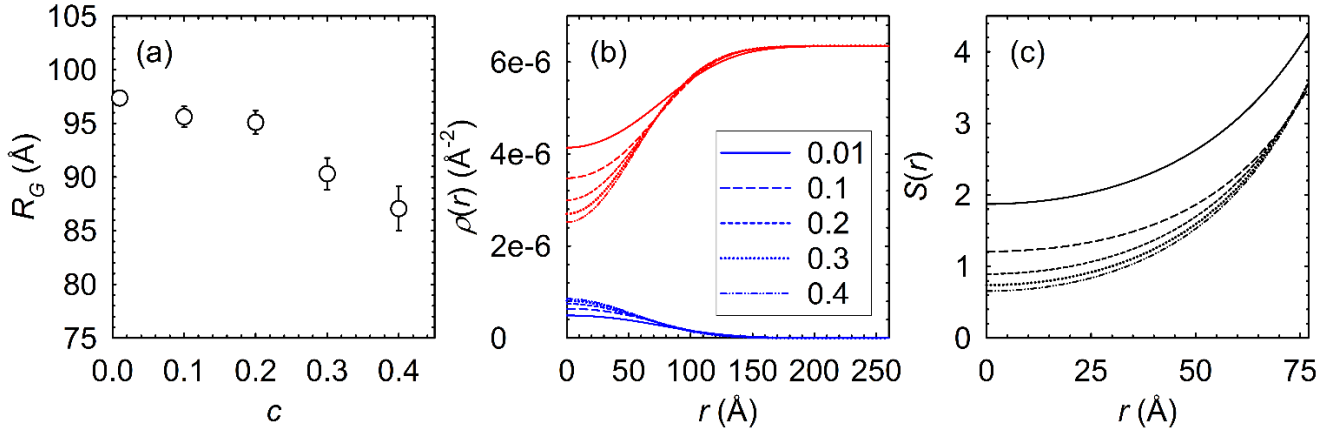

**Figure 4.** The conformational properties obtained from SANS model fitting: (a) $R_G$ as a function of star concentration $c$ in THF solutions; (b) The distributions of THF solvent (red curves) and PS components (blue curves) within a star at different $c$; (c) $S(r)$, the ratio of the volume fraction of THF to that of PS along the radial direction of star $r$.

In conclusion, a close correlation between the molecular solvation and star conformation is revealed using SANS experiments. We provide the first experimental evidence of the detailed conformational dependence of a star polymer on concentration by combining isotopic substitution of the interior region of a star and isotopic contrast in the solvent. The progressive localization of star branches is triggered by the diminishing local solvent density and enhancing excluded volume effect. This conformational variation at the microscopic level is reflected by a size shrinkage at a mesoscopic length scale. We find an analogy between our observed mechanism, which controls the concentration-triggered conformational evolution, and that of the temperature-triggered hydrophobic collapse of a polymer in solution. This study suggests that the influence of solvent fluctuations on the structural variations may be a general property of soft matter systems.

Research presented in this work is supported by the U.S. Department of Energy, Basic Energy Sciences, Materials Sciences and Energy Division. The EQSANS experiment at Oak Ridge National Laboratory's Spallation Neutron Source is supported by the Scientific User Facilities Division, Office of Basic Energy Sciences, U.S. Department of Energy. Synthesis of PS stars used in this research was conducted at the Center for Nanophase Materials Sciences, Oak Ridge National Laboratory, was sponsored by the Scientific User Facilities Division, Office of Basic Energy Sciences, U.S. Department of Energy. We acknowledge the support of the National Institute of Standards and Technology, U.S. Department of Commerce, in providing the neutron research facilities used in this work. We also greatly appreciate the support of SANS beamtime from ILL France and HANARO Korea.

**Associated Content**

1. Synthesis of Polystyrene Star; 2. Small Angle Neutron Scattering (SANS) Experiment; 3. SANS Data Analysis. This material is available free of charge via the Internet at http://pubs.acs.org.

# Influence of Molecular Solvation on Conformation of Star Polymer

Xin Li, Lionel Porcar, Luis E. Sánchez-Diáz, Changwoo Do, Yun Liu, Tae-Hwan Kim, Gregory S. Smith, William A. Hamilton, Kunlun Hong, and Wei-Ren Chen

**SI. Synthesis of Polystyrene Star**

Both fully protonated and partially dueterated polystyrene (PS) stars were synthesized at the Center for Nanophase Materials Sciences (CNMS), Oak Ridge National Laboratory (ORNL).

Polystyrene (PS) star was synthesized using high vacuum with break-seals techniques. [S1]. All reagents were purified by standard procedures deemed for anionic polymerization high-vacuum techniques. [S2] The procedure of synthesis is schematically given in Scheme 1. The linear PS precursor was prepared by polymerization of styrene initiated with sec-butyllithium in benzene at room temperature. After 24 h, the star formation was initiated by adding purified divinylbenzene (DVB) to the solution of living PS. The number of the arms was controlled by the amount of DVB that was added ([DVB]/[Li]). The color of the solution changed from orange/red to darker orange/red as the DVB was added. The reactor was then stirred at room temperature for 120 h and aliquots were taken periodically to monitor the reaction progress. The reaction was terminated with degassed methanol. Solvent/non-solvent fractionation was used to separate the star from lower molecular weight residual precursor and other impurity. PS star with deuterated core was synthesized similarly using styrene-d8 monomer. The molecular weight of the arms and stars were determined by multi-angle laser light scattering in toluene at room temperature. From the Zimm plot given in Figure S1, the molecular weight is found to be 538k. The number of arms calculated from the molecular weight is found to be around 26.

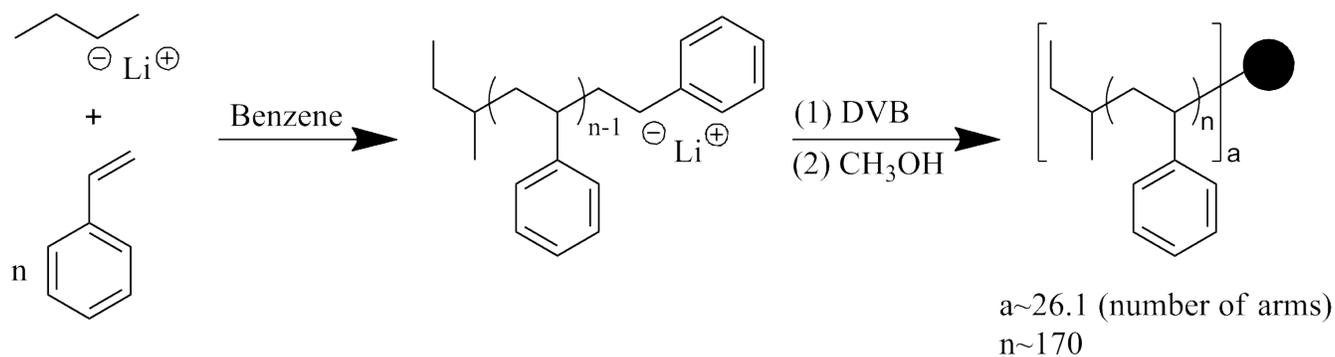

**Scheme S1.** Synthesis route for partially dueterated polystyrene star.

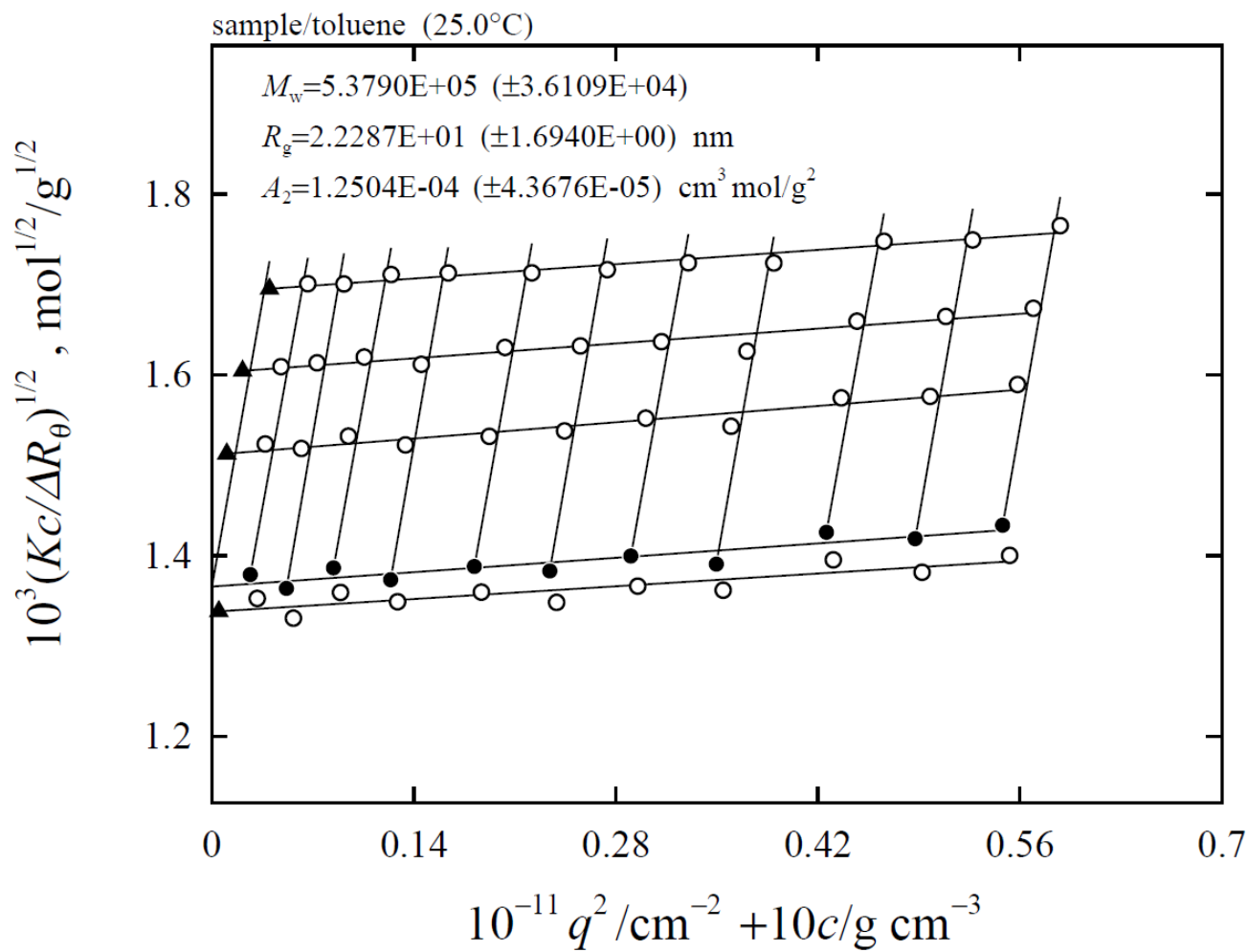

**Figure S1.** The Zimm Plot of the PS star in THF solutions.

## SII. Small Angle Neutron Scattering (SANS) Experiment

Small-angle neutron scattering (SANS) measurements were performed using the D22 SANS spectrometer at the ILL, the NG7 SANS spectrometer at NCNR NIST, the SANS spectrometer at HANARO and the EQ-SANS instrument at SNS ORNL. For the SANS experiment at ILL, NCNR and HANARO, The wavelength of the incident neutron beam was chosen to be 6.0 Å, with a wavelength spread $\Delta\lambda/\lambda$ of 10%, to cover values of the scattering wave vector $Q$ ranging from 0.01 to 0.45 Å$^{-1}$. The measured intensity $I(Q)$ was corrected for detector background and sensitivity and for the scattering contribution from the empty cell and placed on an absolute scale using a direct beam measurement.

For the SANS experiment at SNS, Three configurations of sample-to-detector distance of 4 m, 2.5m, and 1.3m with two neutron wavelength bands were used to cover the $Q$ range of 0.01 Å$^{-1}$ < $Q$ < 0.6 Å$^{-1}$ where $Q = (4\pi/\lambda) \sin(\theta/2)$ is the magnitude of the scattering vector, and $\theta$ is the scattering angle. The measured scattering intensity was corrected for detector sensitivity and the background from the empty cell and placed on an absolute scale using a calibrated standard.

All the SANS measurements were carried out using Hellma quartz cells of 1 mm path length at 25.0 ± 0.1 °C.

## SIII. SANS Data Analysis

### SIII.A Zero-Angle Scattering of Star

The experimental zero-angle scattering $P(0)$ contains the scattering power of a single star. Mathematically, it can be expressed as

$$P(0) = nb_{total}^2 \tag{S2.1}$$

where $n$ is the number density of star polymers in solution. Experimentally, $P(0)$ is obtained by extrapolating the data to the low $Q$ limit using Guinier plot. Given the excess scattering length density (SLD) distribution of a star polymer, $\rho(r)$, $b_{total}$ can be expressed by the following equation

$$b_{total} = 4\pi \int_0^\infty \rho(r) r^2 dr = 4\pi \int_0^\infty \left[ \rho_{polymer}(r) + \rho_{solvent}(v_{THF}(r) - 1) \right] r^2 dr \tag{S2.2}$$

where $\rho_{polymer}(r)$ is the distribution of polymer within a star along the radial direction $r$. $v_{THF}(r)$ is the distribution of THF within a star along the radial direction $r$. And

$$\rho_{solvent} = \gamma\rho_{d-THF} + (1-\gamma)\rho_{h-THF} \tag{S2.3}$$

where $\gamma$ is the deuterium-hydrogen ratio (D/H) of THF, $\rho_{d-THF}$ is the SLD of fully deuterated THF, and $\rho_{h-THF}$ is the SLD of fully protonated THF. Experimentally, $\gamma$ can be precisely controlled by mixing protonated and dueterated THF according to the specified ratio. The total bound scattering length of the polymer components of a star, $b_{polymer}$, is given by

$$b_{polymer} = 4\pi\int_0^\infty \rho_{polymer}(r)r^2 dr \tag{S2.4}$$

If we define

$$V_{es} \equiv 4\pi\int_0^\infty (1 - v_{THF}(r))r^2 dr \tag{S2.5}$$

then Eqn. (S2.2) becomes

$$\begin{aligned} b_{total} &= b_{polymer} - \rho_{solvent}V_{es} \\ &= -(\rho_{d-THF} - \rho_{h-THF})V_{es}\gamma + b_{polymer} - \rho_{h-THF}V_{es} \end{aligned} \tag{S2.6}$$

If we define

$$\Delta b_s = (\rho_{d-THF} - \rho_{h-THF})V_{es} \tag{S2.7}$$

and

$$b_0 = b_{polymer} - \rho_{h-THF}V_{es} \tag{S2.8}$$

then Eqn. (S1) becomes

$$P(0) = n(b_0 - \Delta b_s \gamma)^2 \tag{S2.9}$$

and

$$\frac{dP(0)}{d\gamma} = 2n\Delta b_s(\Delta b_s\gamma - b_0) \tag{S2.10}$$

For the PS stars, depending on concentration, $V_{es}$ is found to be on the order of $10^7 \text{Å}^3$, and $\Delta b_s$ is on the order of 10Å. Since $b_{polymer}$ ~1Å for a fully protonated star and ~4Å for a partially deuterated star, $b_0$ is found to be on the order of 0.1Å for a fully protonated star and 1Å for a partially deuterated star. As a result, $dP(0)/d\gamma$ stays positive for fully protonated stars at all concentrations studied, and $dP(0)/d\gamma$ shows a transition from negative to positive values for partially deuterated stars with increasing concentration.

**SIII.B Model of Coherent Scattering Cross Section of Star**

To extract the quantitative conformational information from an experiment, a model of scattering function is proposed for SANS data analysis based on the aforementioned conformational picture of a star polymer. Because the density of a star polymer exhibits the highest value at its molecular center, the intra-molecular SLD profile of a PS star is modeled by a convolution of a step function and normal distribution. It should be mentioned that the DVB core is treated as a point in the center of the star polymer, with all PS arms connected to it. The mass fraction of the DVB core is 14.2%, calculated from the chemical characterization methods. From the chemical synthesis procedures we know that the DVB core is highly cross-linked, and its packing density is much higher than the PS arms. It is also known that the scattering intensity is proportional to the sixth power of the particle size. Therefore, the contribution from the DVB core is neglected in our model, and the quality of the SANS data fitting proves that our model has caught the main feature in the conformation of a PS star. Namely

$$\rho_{PS}(r) = \Pi(r) \otimes N(0, \sigma^2) \tag{S2.11}$$

where the step function $\Pi(r)$ is

$$\Pi(r) = \begin{cases} 1 & r < R \\ 0 & r > R \end{cases} \tag{S2.12}$$

$N(0, \sigma^2)$ is the normal distribution with mean of 0 and variance of $\sigma^2$. It takes the following expression

$$N(0, \sigma^2) = \frac{1}{\sqrt{2\pi}\sigma} \exp\left(-\frac{r^2}{2\sigma^2}\right)$$

(S2.13)

The polydispersity of the star polymer has been taken into account in this normal using its standard deviation $\sigma$. With larger polydispersity, the normal distribution Eqn. (S2.13) will have a larger standard deviation $\sigma$, and the features in Figure 3, panel (b) in the manuscript will become more smeared. It is important to point out that the $\rho_{PS}(r)$ for a PS star in THF solution is collectively determined by the intra-star mass distributions of both its polymer component and the invasive THF. From Eqn. (S2.11) it can be shown that the volume distribution function of polystyrene, $\phi_{PS}(r)$, takes the following expression

$$\phi_{PS}(r) = \frac{\phi(0)}{2}\left[erf\left(\frac{R-r}{\sigma}\right) + erf\left(\frac{r+R}{\sigma}\right)\right] \tag{S2.14}$$

Again the decay of the intra-molecular PS volume fraction is approximated by the reverse $S$ curve given by the error function on the right hand side of Eqn. (S2.14). For general soft colloidal systems such as star polymers, significant conformational variation occurs when the concentration is above the overlap concentration $c^*$. This is a result of the increasing congestion and interpenetration between the neighboring PS stars due to the inter-star interaction. To investigate the conformational evolution of PS star as a function of its concentration, we assume that qualitatively the conformation of PS star can still be described by the dense-core profile given in Eqn. (S2.11). The effect of concentration only changes the quantitative values of $R$ and $\sigma$ in Eqns. (S2.11) and (S2.14). Assuming the isotopic effect on the volume distribution of PS can be ignored, for a partially deuterated PS star consisting of a deuterated core and a protonated periphery, the corresponding volume distribution function of polystyrene, $\phi_{PS-pd}(r)$, can be expressed as

$$\phi_{PS-pd}(r) = \phi_{PS}(r) = \phi_{PS-D}(r) + \phi_{PS-H}(r) = \frac{\phi(0)}{2}\left[erf\left(\frac{R-r}{\sigma}\right) + erf\left(\frac{r+R}{\sigma}\right)\right] \tag{S2.15}$$

where $\phi_{PS-D}(r)$ and $\phi_{PS-H}(r)$ represent the volume distributions of the deuterated and the protonated PS in a partially deuterated PS star, respectively. Since the density distribution of the inner deuterated region of the PS star also follows the expression given by Eqn. (S2.11), $\phi_{PS-D}(r)$ can be further expressed as

$$\phi_{PS-D}(r) = \frac{\phi(0)}{2}\left[erf\left(\frac{R_D-r}{\sigma_D}\right) + erf\left(\frac{r+R_D}{\sigma_D}\right)\right] \tag{S2.16}$$

where $R_D$ and $\sigma_D$ are the parameters quantifying the conformation of the deuterated inner region. The following equation must be satisfied assuming incompressibility,

$$\phi_{PS-D}(r) + \phi_{PS-H}(r) + \phi_{THF}(r) = 1 \tag{S2.17}$$

where $\phi_{THF}(r)$ gives the average volume distribution of the invasive THF along the radial direction. Total mass conservation requires the following two equations to be followed:

$$4\pi \int_0^\infty [\phi_{PS-D}(r) + \phi_{PS-H}(r)] r^2 dr = constA$$
$$4\pi \int_0^\infty \phi_{PS-D}(r) r^2 dr = constB \tag{S2.18}$$

The two constants, *constA* and *constB*, are the total volume of the polymeric part and the deuterated part, respectively. They are obtained from fitting the dilute case and fixing them for the concentrated solutions. Eqn. (S2.18) serves as the constraint to reduce the number of free parameters by 2 and provide better precision for the curve fitting. From Eqns. (S2.15) – (S2.17), the SLD for a fully protonated PS star polymer, $\rho_{PS-h8}(r)$, can be expressed as

$$\rho_{PS-h8}(r) = \rho_{PS-H}[\phi_{PS-D}(r) + \phi_{PS-H}(r)] + \rho_\gamma \phi_{THF}(r) \tag{S2.19}$$

where $\rho_{PS-H}$ is the SLD of the protonated polymeric components of PS, and $\rho_\gamma$ is the SLD of bulk THF at a certain D/H fraction, $\gamma$, which can be expressed as

$$\rho_\gamma = \gamma \rho_{THF-D} + (1-\gamma) \rho_{THF-H} \tag{S2.20}$$

$\rho_{THF-D}$ and $\rho_{THF-H}$ represent the SLD of deuterated and protonated THF, respectively. Therefore, the excess SLD for fully protonated star polymers, $\Delta\rho_{PS-h8}(r)$, is found to be

$$\Delta\rho_{PS-h8}(r) = \rho_{PS-H}[\phi_{PS-D}(r) + \phi_{PS-H}(r)] + \rho_\gamma \phi_{THF}(r) - \rho_\gamma \tag{S2.21}$$

Similar to Eqns. (S2.18) and (S2.20), the SLD and excess SLD for partially deuterated PS star polymers are found to be

$$\rho_{PS-pd}(r) = \rho_{PS-H}\phi_{PS-H}(r) + \rho_{PS-D}\phi_{PS-D}(r) + \rho_\gamma \phi_{THF}(r) \tag{S2.22}$$

and

$$\Delta\rho_{PS-pd}(r) = \rho_{PS-H}\phi_{PS-H}(r) + \rho_{PS-D}\phi_{PS-D}(r) + \rho_\gamma \phi_{THF}(r) - \rho_\gamma \tag{S2.23}$$

where $\rho_{PS-D}$ is the SLD of the deuterated polymeric components of PS. Based on this model, the parameters $R$ quantifying the size of the hard-core and $\sigma$ measuring the peripheral fuzzy region can be obtained from SANS data analysis. The scattering amplitude $F(Q)$ of SANS is given by the following equation:

$$F(Q) = \int \Delta \rho(r) \exp(-iQ \cdot r) d^3 r \tag{S2.24}$$

It is worth mentioning that, due to the confinement effect, the structure of intra-star invasive THF is expected to be different from that of bulk THF. This is indeed reflected in Eqns (S2.20) and (S2.22). The contribution from the invasive THF to the coherent scattering is represented by the difference between $\rho_r \phi_{THF}(r)$ and $\rho_r$. The form factor $P(Q)$ used in the data analysis can be obtained from

$$P(Q) = |F(Q)|^2 \tag{S2.25}$$

The measured coherent scattering cross section $I(Q)$ of PS star polymer solutions can be expressed as

$$I(Q) = n_p P(Q) S(Q) \tag{S2.26}$$

Where $n_p$ is the number density of PS star polymers. $S(Q)$ is the structure factor which contains the information of spatial arrangement of PS stars in solution.

**SIII.C Results of SANS Data Analysis**

The SANS coherent scattering cross sections $I(Q)$ of 1 wt% THF solutions of fully protonated PS stars (PS-h8) and partially deuterated stars (PS-pd) are given respectively in Figure S2a and Figure S2b.

In the dilute limit the value of $S(Q)$ can be approximated as 1. Therefore according to Eqn. (S2.25) the only structural information contained in the measured $I(Q)$ is the form factor $P(Q)$. One obvious difference between the evolution of $I(Q)$ for the reference PS-h8 and that of PS-pd is immediately noticed. Upon changing the D/H fraction in THF from 1 to 0.6, a steady decrease in the magnitude of $I(Q)$ for PS-h8 solutions is observed. Meanwhile, the absolute value of $I(Q)$ for PS-pd is significantly less than that for PS-h8 at any given D/H ratio of THF. As an example, a nearly two orders of magnitude difference between their corresponding $I(Q)$ is observed around $Q = 0.01$ Å$^{-1}$ when the D/H ratio is 100/0. It can be understood that this observed diminishing scattering contrast is the consequence of the partially deuterated molecular structure of PS-pd. However, contrary to the evolution of $I(Q)$ for PS-h8, $I(Q)$ for PS-pd is seen to increase significantly with a decrease in the D/H ratio of THF. Moreover, it is also found that the $I(Q)$ for PS-pd cannot be completely matched within this range of D/H ratio. These two observations present clear evidence that the relative intra-molecular spatial arrangement between their protonated and deuterated components is indeed characterized by a regular pattern universal for every PS-pd star. The details of this conformational feature is further revealed by the radial SLD distributions $\rho(r)$ given in Figures S1(c) and S1(d). The reducing scattering contrast due to the decrease in the D/H ratio of THF results in a steady decrease in $\rho_{PS-h8}(r)$ as shown in Figure S2(c), as well as that of $\Delta\rho_{PS-h8}(r)$ according to Eqn. (S2.18). The reducing $\Delta\rho_{PS-h8}(r)$ in turn renders a steady decrease in the $I(Q)$ for PS-h8 shown in Figure S2(a). On the other hand, the characteristic variation of the $I(Q)$ for PS-pd star (Figure S2(b)) induced by the varying D/H ratio is a reflection of its specifically tailored molecular architecture.

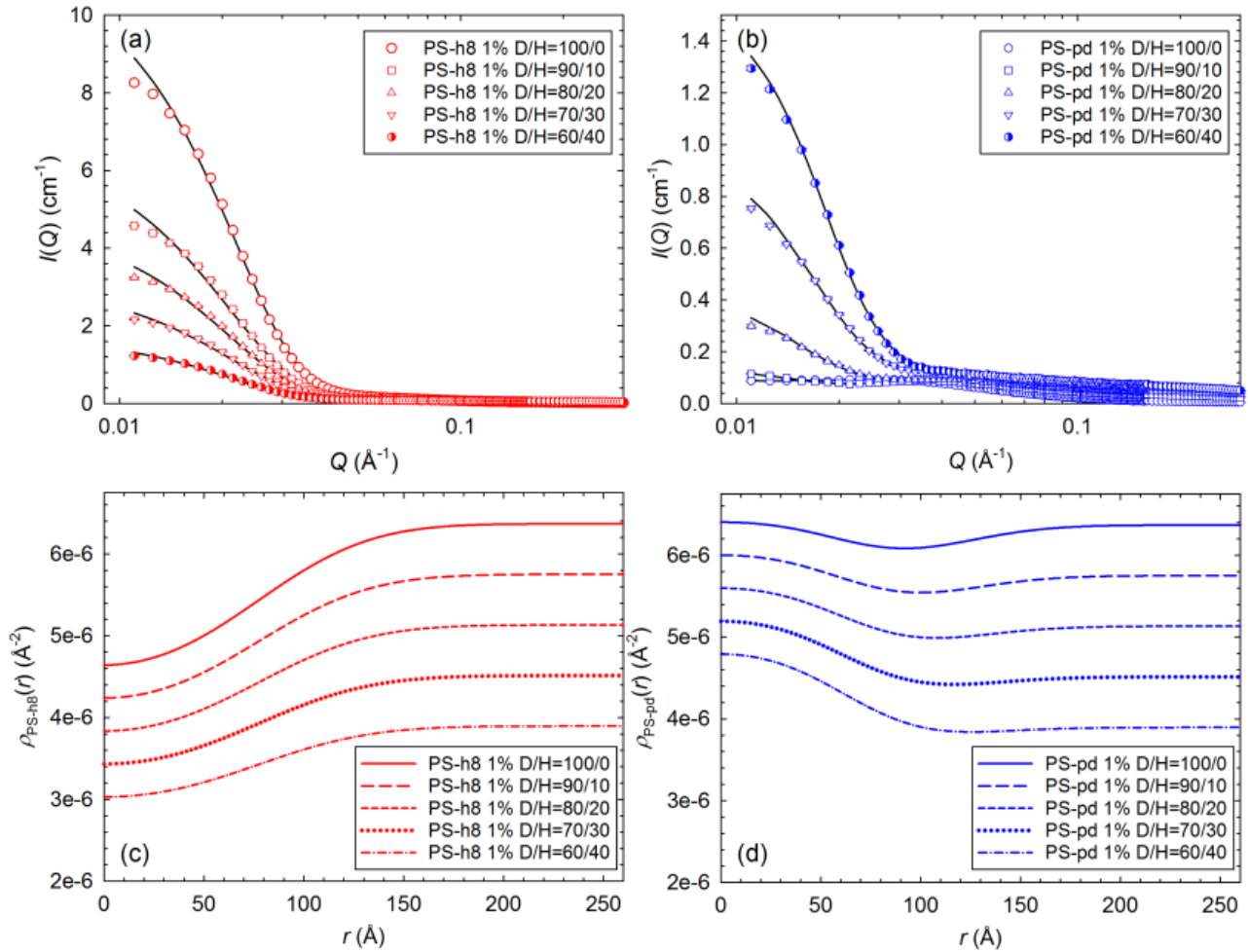

**Figure S2.** The absolute SANS intensity I(*Q*) (red and blue symbols) and fitting curves (black curves) of fully protonated PS star (a) and partially deuterated PS star (b) as a function of the D/H ratio of the THF solutions. The concentration of PS star in the THF solutions is 1 wt%; therefore, the coherent scattering contribution from inter-star spatial arrangement can be ignored. The corresponding radial SLD distributions $\rho(r)$ of a single star are given in (c) and (d).

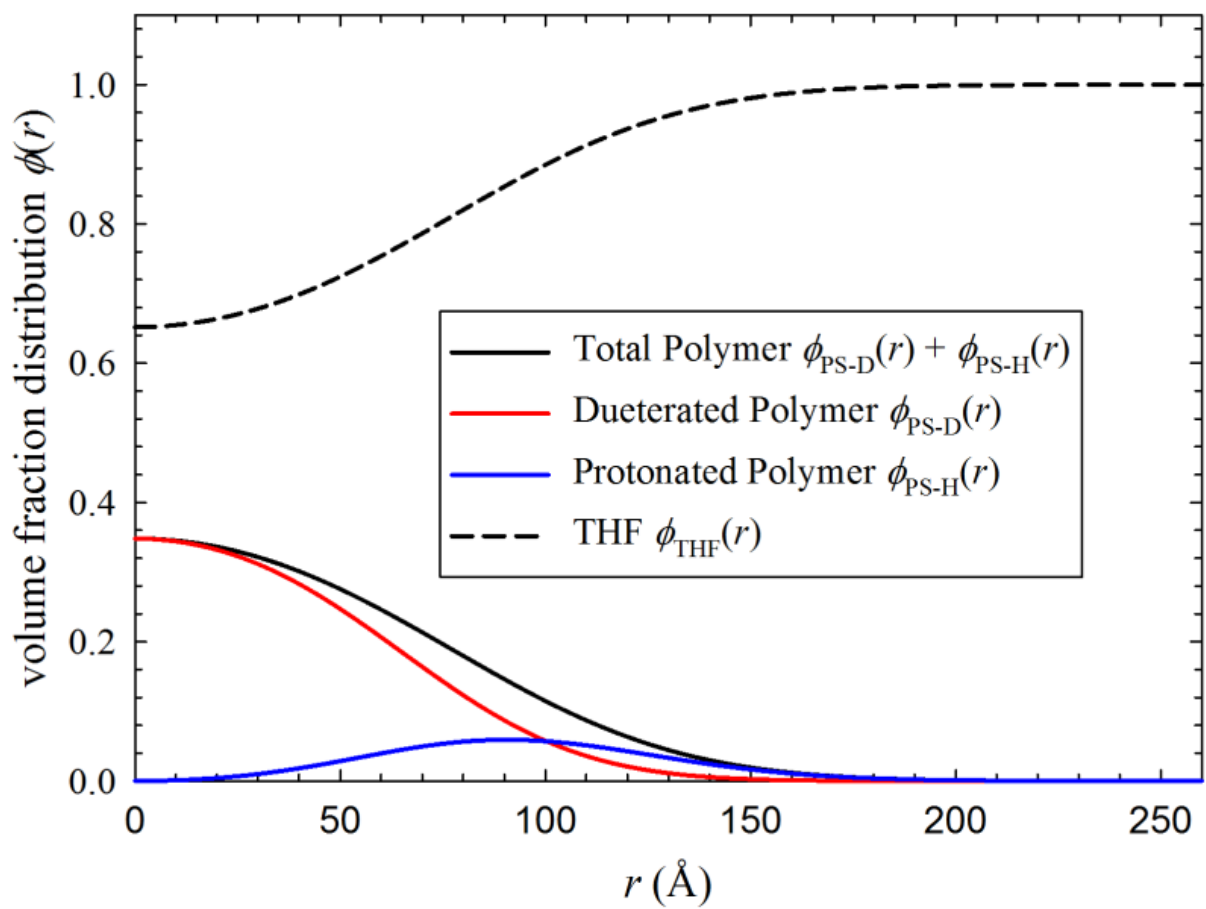

**Figure S3.** The volume fraction distribution $\phi(r)$ of each constitutive component of partially deuterated PS star suspended in THF solvent. The solid black, blue, red, and dashed lines are the $\phi(r)$ for the total polymer, deuterated polymer, protonated polymer, and invasive THF, respectively.

Figure S3 present the volume fraction distribution $\phi(r)$ of each constitutive component of PS-pd star obtained from SANS data analysis. It is clearly seen that majority of the deuterated polymer ($\phi_{PS-D}(r)$, red curve) remains in the molecular interior while the molecular periphery is occupied mostly by the protonated component ($\phi_{PS-H}(r)$, blue curve). The total polymer volume fraction distribution ($\phi_{PS-D}(r)$ + $\phi_{PS-H}(r)$, black curve) is characterized by the dense-core profile. The volume distribution function of THF, $\phi_{THF}(r)$, is given by the dashed line. This radial distribution implicitly suggests that, due to the effect of confinement, the packing of invasive THF is different from that of THF at its bulk state. When the D/H ratio of THF is 100/0, the coherent scattering is mainly contributed by the protonated outer layer because the SLD of the deuterated molecular interior is close to that of the fully deuterated THF.

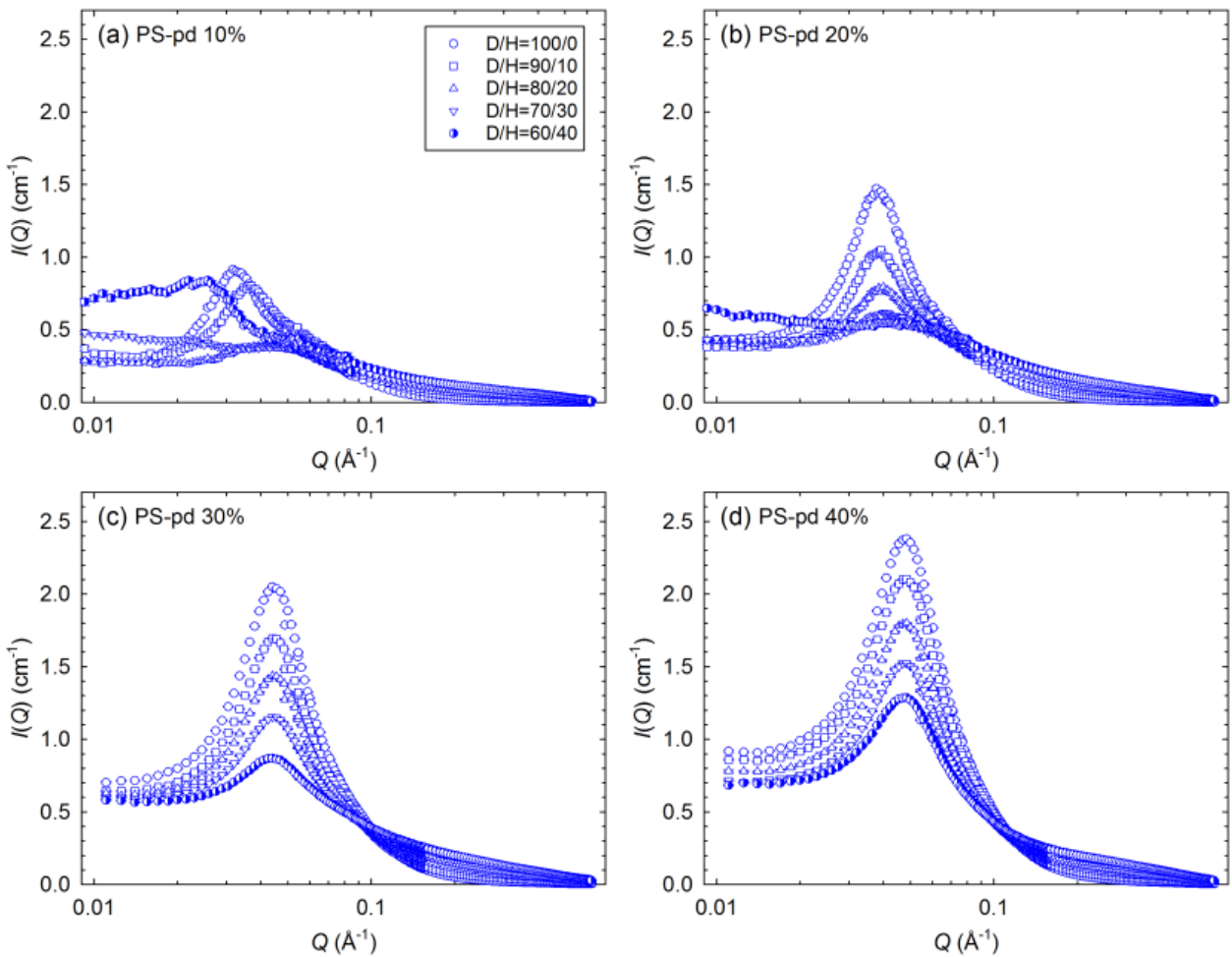

**Figure S4.** The absolute SANS intensity $I(Q)$ collected from the THF solutions of partially deuterated PS stars (PS-pd) at different concentrations.

Therefore, $\Delta\rho_{PS-pd}(r)$ is characterized by a profile which exhibits its maximum near the peripheral region as indicated by Figure S2(d). Further changing the D/H ratio from 100/0 to 60/40, the difference between the SLD of the protonated periphery and that of THF is diminishing due to the increasing amount of protons in THF. Meanwhile, the difference between the SLD of the deuterated interior and that of THF is progressively widening for the same reason. As a result, the $\Delta\rho_{PS\text{-}pd}(r)$ gradually evolves into the dense-core profile which manifests the dominance of coherent scattering contributed by the deuterated interior. This observation suggests that this chemical substitution does not fundamentally change the intra-molecular structure. However, indeed it renders a characteristic intra-molecular distribution of scattering length. Contrary to the $\Delta\rho_{PS\text{-}h8}(r)$ characterized by the dense-core picture, the evolution of $\Delta\rho_{PS\text{-}pd}(r)$ exhibits a non-monotonic dependence on the D/H ratio of THF. It can be seen from Eqn. (S2.22) that this unique feature, along with $\rho_n$, which is adjusted by varying the D/H ratio of THF, allows us to determine the conformational evolution as a function of concentration.

In Figures S4 we present the SANS $I(Q)$ of PS-pd obtained at different concentrations $c$. As we have qualitatively demonstrated in Figure 1 of this report, the inter-star interpenetration indeed enhances the scattering contrast. As a result, upon increasing $c$, the intensity of the $I(Q)$ for PS-pd is seen to increase continuously. This observation presents a stark contrast in comparison to the diminishing $I(Q)$ for PS-h8 given in Figure S5.

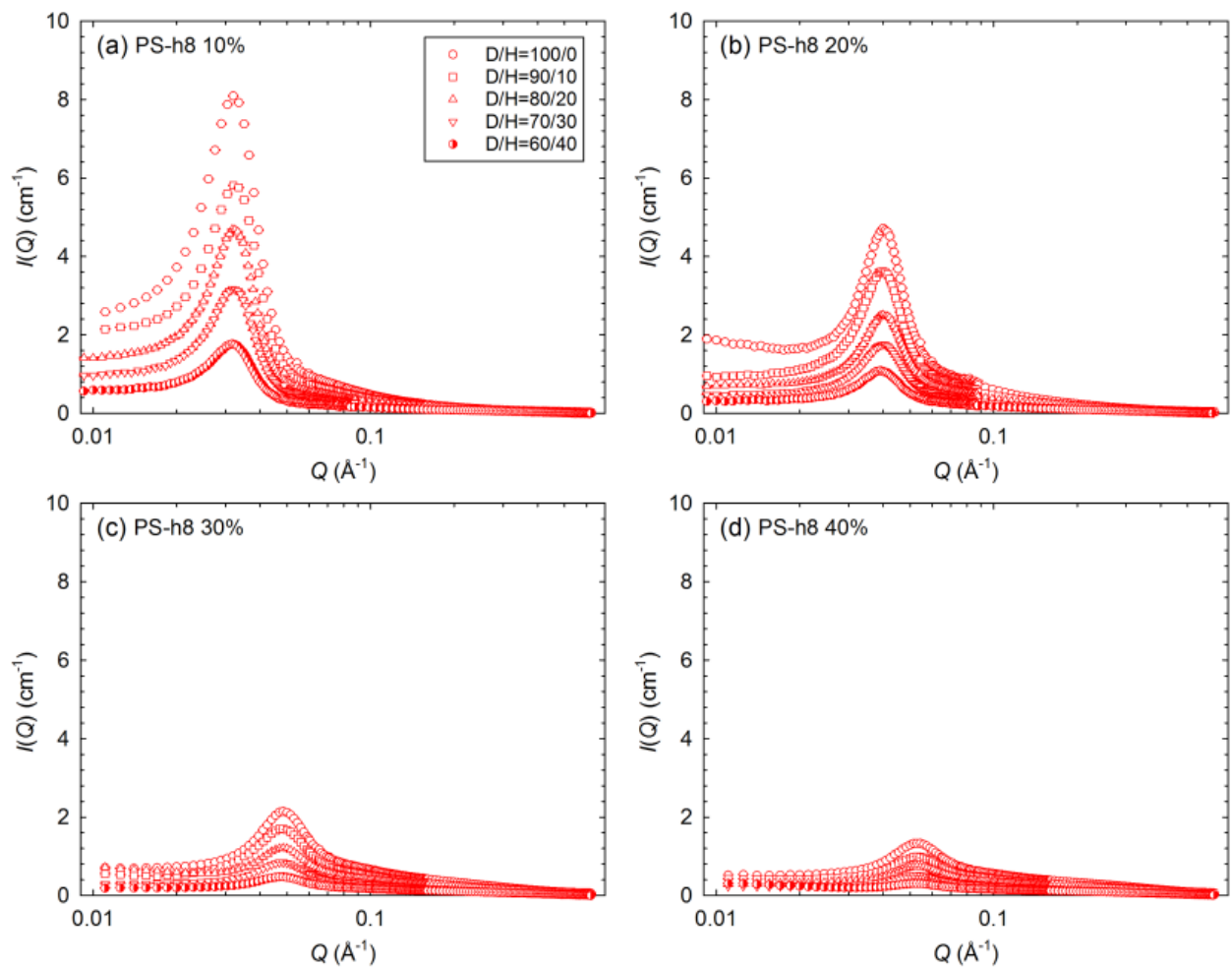

**Figure S5.** The absolute SANS intensity $I(Q)$ collected from the THF solutions of fully protonated PS stars at different concentrations.

The SANS $I(Q)$ presented in Figures S4 and S5 not only contains the conformational information of a single star in the form factor $P(Q)$ but also that of the inter-star spatial arrangement in the structure factor $S(Q)$. Since $I(Q)$ can be factorized as the product of $P(Q)$ and $S(Q)$, at a certain $c$ the former can be effectively compartmentalized by normalizing the $I(Q)$ obtained at a given $\gamma$ to that obtained at $\gamma =$ 100/0. While the intra-star mass distribution for both PS-h8 and PS-pd can be satisfactorily described by the dense-core picture, the SLD profile of PS-pd is characterized by a core-shell distribution as shown in Figure S2(d) because of isotopic labeling. This non-monotonic SLD profile indeed provides a unique feature which greatly facilitates the SANS model fitting. Figure S6 presents the theoretical form factors $P(Q)$ for PS-h8 (Figure S6a) and PS-pd (Figure S6b) calculated based upon their respective SLD profiles. Upon changing $\gamma$, although the intensity of $P(Q)$ for PS-h8 keeps evolving, the locations of the first minimum and the first maximum remain invariant. Meanwhile, for PS-pd, both the intensity and the minimum and maximum positions of $P(Q)$ are seen to change. For the sake of clarity, all these theoretically calculated form factors are normalized to the total scattering power, $P(Q=0)$, of the PS-h8 star polymer solution in $D_2O$.

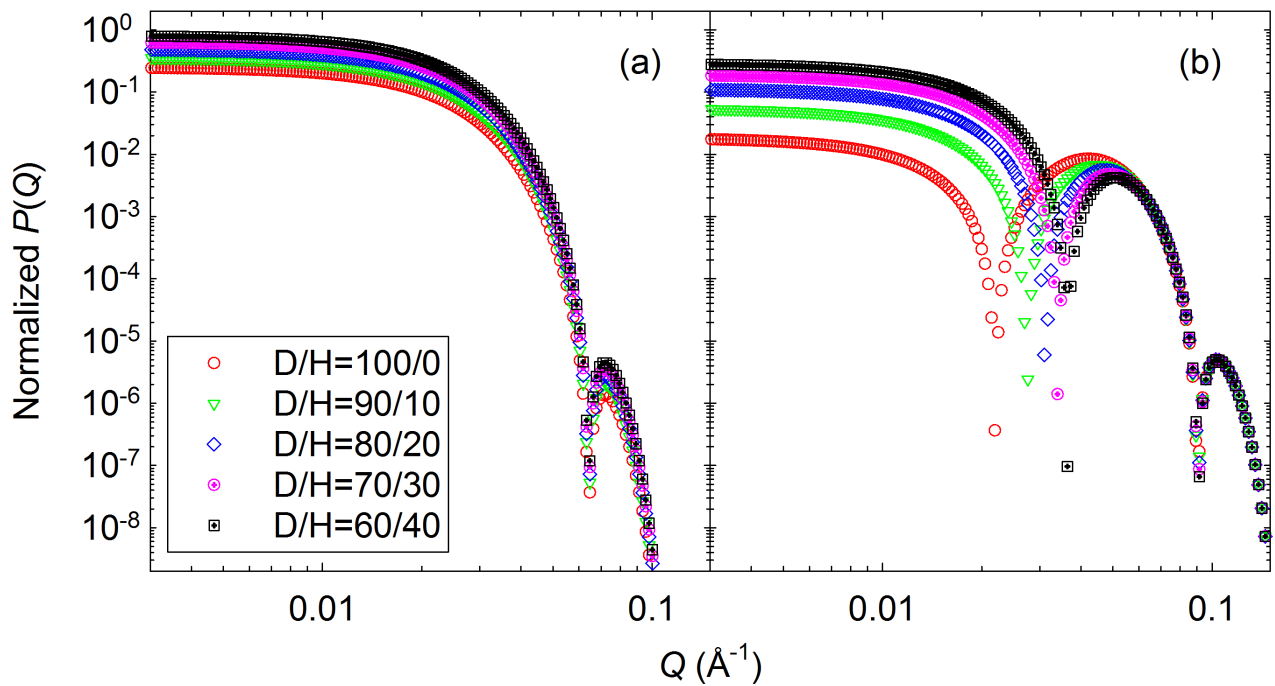

**Figure S6.** Theoretical form factors $P(Q)$ for PS-h8 (panel a) and PS-pd (panel b).

As shown in Eqn. (S2.26), the normalization by dividing two scattering functions at different contrasts, cancels $n_p$ and the structure factor $S(Q)$, giving the same result as dividing the two form factors at those contrasts. It keeps the conformational evolution of a single star under the influence of concentration. It should be mentioned that in the presented normalization of the measured $I(Q)$s, the contribution of the blob scattering of the star polymers and the instrument resolution are involved, which makes it not directly the same as dividing two curves in Figure S5 (a) or (b), respectively. Considering both the clarity of the difference between two contrasts and data statistics, the normalization of the $I(Q)$ at D/H=80/20 to D/H=100/0 ($I_{80/20}(Q)/ I_{100/0}(Q)$) is selected for fitting and presentation.. As demonstrated in the main manuscript, for fully protonated star solutions, the normalized $I(Q)$ shown in Figure 3(a) is rather featureless. On the contrary, upon increasing $c$ the normalized $I(Q)$ for the solutions of partially deuterated star, shown in Figure 4(b), is seen to vary characteristically. In the case of the PS-pd star polymer, because of the shifted first minimum and first maximum when varying the contrast, the operation of division gives the minima and bumps as shown in Figure 4(b), which provides the possibility to reveal the conformational information through fitting. To obtain reliable conformational properties, all the SANS results obtained at different $\gamma$ are used in our model fitting proposed in Section SIIIB via a universal optimization approach. These detailed features, such as the locations of minima, prove critical for identifying the universal minima in the space of conformational parameters of our model. We are able to determine the distributions of solvent and polymer within a star separately and accurately.